# Collecting and Preserving Videogames and Their Related Materials: A Review of Current Practice, Game-Related Archives and Research Projects

**Megan A. Winget & Caitlin Murray**
School of Information, University of Texas at Austin; 1 University Station, D7000, Austin, TX 78712-0390.
megan@ischool.utexas.edu

**Abstract**

**This paper reviews the major methods and theories regarding the preservation of new media artifacts such as videogames, and argues for the importance of collecting and coming to a better understanding of videogame "artifacts of creation," which will help build a more detailed understanding of the essential qualities of these culturally significant artifacts. We will also review the major videogame collections in the United States, Europe and Japan to give an idea of the current state of videogame archives, and argue for a fuller, more comprehensive coverage of these materials in institutional repositories.**

**Videogame Preservation & The Creation Process**

The videogame industry is at a critical moment in its history. As videogames are increasingly recognized as important cultural artifacts, the games are becoming more and more difficult to access and play, videogame pioneers are getting older and older, and their primary materials are being thrown away as companies go out of business, or are deteriorating in garages and attics across the nation. The desire to preserve and protect this material and intellectual culture is growing, as is the need to provide primary source material for the study and advancement of the industry. As game developer Warren Spector notes,

> "The fact is that the history of the videogame business is being written every day and, sadly, being lost just about as quickly. So here's my thought: We need to step up NOW and recognize the cultural and academic importance of videogames. Luckily, we're a young enough medium that nearly all of our serious practitioners are still alive and available to be interviewed and/or solicited for contributions to an archive of research and reference materials." (Spector, 2007)

Preserving and providing meaningful access to any traditional artistic medium is based on hundreds of years of knowledge and practice. Initially through the process of trial and error, and then through scientific/chemical principles, people have been conserving traditional art forms for hundreds of years. New media artifacts like videogames do not have that history of production and scholarship, nor is there time to "hope for the best," in terms of preservation. For example, while it is still possible to look at unpreserved five-hundred-year-old paintings and sculptures, and in many cases, we can still look at preliminary studies and drawings for those works – there is a significant risk of losing a new media artifact as soon as *ten years* after its initial creation.

> "We are faced with the potential disappearance of our cultural heritage if we don't act soon and act together to preserve digital materials... We have learned from our experience that long-term preservation of digital content is dependent on influencing decisions of content providers from the moment of creation." ~Laura Campbell, Associate Librarian for Strategic Initiatives at the Library of Congress

Because new media art generally, and videogames in particular, have a significant digital component, they tend to rapidly become, at best, inaccessible; and at worst, irretrievably lost. With funding from the NEH and IMLS, scholars in the related field of new media art have produced numerous theoretical and practical tracts with which to work, including the development of a notation framework for media art (Rinehart, 2004); a systematic review of emulation as a strategy for preservation of a multimedia work (Rothenberg, 2006); and the formulation of agreed upon theories and methods for the preservation of variable media art (Depocas, Ippolito, & Jones, 2003).

These projects focus, to one degree or another, on the end product. However, in the archival literature, there is a growing recognition that "preservation [of digital artifacts] begins with creation" (Waters & Garrett, 1996). Before we can attempt to develop a robust collection and preservation model, or to make suggestions on enduring creation techniques, the colleting community needs to have a better idea of how new media is currently created. At the University of Texas at Austin's School of Information, in an attempt to create a model of "typical" videogame creation, we are initiating a research project called *[re]Create* to inventory the artifacts of the videogame creation process, and document the artistic methods of inspiration within different contexts of creation. We believe that this information will be significant for the videogame industry as well as the more general community interested in new media art and science; and it will be valuable not only for those interested in relevant representation and preservation of these materials, like conservators, archivists, and librarians; but collectors, curators, and scholars will also find this information useful as a reference to new media creativity conditions and value. Research questions for this project include: How is the game creation process managed? What are the steps / processes involved in video game production? What kinds of people are involved at each step of the creation process? What kinds of artifacts are created in the early / middle / late stages of the creation lifecycle? What is the process of their creation, and how do those artifacts then feed into the creation process? Is there any industry-wide recognition of the value of these creation artifacts? Are these artifacts being preserved? If so, how and by whom?

Instead of focusing on the end product, the [re]Create will focus on the creative process, interviewing videogame artists, programmers, producers, and developers to get a better conception of the creative process, and developing initial inventory lists of the collected artifacts of creation owned and stored by these individuals and companies. These materials will be made available on the web for scholars, artists, and the general public through the newly developed Videogame Archive, a scholarly research collection at the Center for American History, at the University of Texas at Austin. This project's investigators believe that better understanding of this creative process will enable better and more relevant collection, access, and preservation strategies for all types of collecting institutions that are interested in these relevant and important cultural artifacts.

### *Current Preservation Methods: A Critical Review*

The variable media art community, which includes the videogame industry, currently utilizes four digital preservation strategies, all focused on the end product. The first three methods have technical origins, and are based on general digital preservation practices. Related to "the viewing problem," they are: *refreshing*, the upgrade of storage mechanisms; *migration*, the premeditated upgrade of file formats; and *emulation*, which focuses on development of Ur-operating systems able to run obsolete media. The fourth option, developed by and for the new media art community, is *re-interpretation* (Depocas et al., 2003); a method intimately related to the presentation, exhibition, and performance of an interactive variable media art object.

### **The Viewing Problem**

The primary problem with the longevity of digital documents is the "viewing problem" (Besser, 2000). Unlike analog or physical information, which tends to exist independent of human involvement, digital information needs constant intervention in order to survive. History has shown that digital documents are problematic by default. Whereas we can actually look at the Sistine Ceiling, created five hundred years ago, or play games, like *go* invented over a thousand years ago; it is difficult if not impossible to view simple documents on 8-inch floppy disks created in the last twenty years, even if there has been an immediate, proactive role in preserving them. Without concerted effort on the part of archivists and preservationists, digital objects quickly become obsolete or inaccessible due to unforeseen, although anticipated, advances in information technology.

*Migration* and *emulation* are the two primary methods in managing the problem of obsolete file formats (Waters & Garrett, 1996). *Migration* focuses on the files themselves, periodically updating files in new software formats. With migration, it quickly becomes a question of whether the conservation/preservation community is trying to preserve access to the physical content of a work, or trying to preserve access to its deeper meaning. It becomes a very sticky business wherein an archivist or curator has to make major artistic choices specifically related to format.

The second method of file format preservation is *emulation*, which can either be at the system or software level. System emulation focuses on developing systems that mimic the hardware used to create or run the original artifact. By writing just a few hardware emulators, dozens of operating systems, thousands of applications, and millions of documents could be accessed (Rothenburg, 1999). Emulation is currently the most widely accepted method of digital preservation and access for variable media art (Rinehart, 2002; Rothenburg, 2003), including videogames. While this is a legitimate method of preserving more simplistic artifacts like text documents, emulation is problematic for more complicated works, or works that convey artistic themes. For example, the processor speed of machines doing the emulation is often significantly faster than that of the original; and because there is no way to fully document the original speed of older videogames, emulated games are often much faster than originally intended. A recent emulation of *Moon Dust*, one of the earliest computer games, was shown to its original designer Jaron Lanier who contended that it was a completely different game than the one he designed because the pacing was different, and he would not claim authorship of this new game (Besser, 2001).

Another major disadvantage of current emulation strategy in relation to videogames is the focus on stand-alone computers rather than the network. Many current videogames, particularly the massively multiplayer online games, integrate and depend on the network for development of the primary work – for example *World of Warcraft* (Blizzard Entertainment, 2005) is an immersive massively multiplayer online (MMO) game that collects data in real time from tens of thousands of users to enable game play itself. Emulating the hardware and software will only go so far in truly preserving this work.

Emulation will never be able to reproduce the entire Internet environment needed for this type of piece, although one might be able to mitigate this problem through collection of very specific information about the work's functionality. Any emulation will necessarily be partial; and the resulting preserved-entity would only be able to serve as a simulation or representation of the original experience, and would therefore serve primarily as a snapshot or fragment to preserve some critical historical evidence of the original. Emulation, in the terms described above, would diminish the immediacy of the gaming experience. Whether that diminution was significant would depend on the presence of a comprehensive representation of that object or experience, able to describe the creator's intent when creating the work.

In videogames, like any autographic art, every difference makes a difference (Goodman, 1976), and because software emulation necessarily disrupts formal elements of a work like pacing, aspect rations, and format, it presents even more profound artistic problems than does system emulation. The related problems of accountability and authenticity lead to the complex issue of defining the nuances in individual pieces. What are the essential elements of a work? Who decides what's important? The creator? The critics? The users? Change is inevitable, particularly in digital media – but is it possible to track those changes? What sort of benchmarks can the preservation community define for individual works that would meaningfully measure change? These questions are not new, and have been plaguing art conservators since the beginning of the profession over a hundred years ago (Glueck, 1980). The dilemma really came to a head in the post-war era, with the dual emergence of "modern" art, which tended to have more ephemeral qualities than the art of the old masters (Wyer, 1988), and "modern" science, which was able to take advantage of new chemical techniques to halt or repair the progress of time on works of art. Specifically, there are two issues in traditional art conservation that are particularly germane to the discussion of videogame preservation: the reliability and/or appropriateness of seeking out creative intent; and the reality of ageing itself – should conservators try to keep the work in pristine condition, or is it more honest to allow for some ageing?

**The Concept of Intent & The Reality of Ageing**

The first problem, that of creative intent, is particularly notable, because much of the current thinking on digital art preservation has an artist questionnaire as one of the first and central means of defense (Ippolito, 2003) (Rinehart, 2002) (Besser, 2001). However, for the last fifty years, conservators have been debating the appropriateness of seeking out artistic intent (Lyas, 1983; Wimsatt & Beardsley, 1948). Comprehension of intent is a very complex process, sometimes not fully understood even by the creator himself (Sloggett, 1998); it is often ancillary to received wisdom about the piece (Dykstra, 1996); and more often than not, conflicts with what a conservator is, or should be, willing to do (van de Wetering, 1989). If archivists, curators, and conservators had a deeper understanding of the general creation behaviors and methods used by new media artists in general, perhaps discussion of intent would become less important to the preservation framework as a whole.

The matter of ageing gracefully is also relevant to the discussion of videogame preservation, although there are considerable differences between physical works, which deteriorate slowly over time; and digital works, which tend to simply disappear or become impossible to access. To make matters even more complicated, because many contemporary videogames depend on massive numbers of people to use the system and make the game interesting, ephemerality and variability are integral characteristics of the work. Should those characteristics be overridden in the name of technical possibilities or aesthetics? Again, the problem of ageing has also been a major issue for traditional art conservators. Picasso and Braque "would rather have had a painting disintegrate than see it undergo plastic surgery" (Richardson, 1983). However, not all people see the problem in those black and white terms, and there is a question of whether the will of those conservators, who define themselves as "restoration minimalists," might take precedence over the creative process of the creator. By refusing to repair decaying art works, even if the technology is available, is the conservator then locking the artwork into a single moment in time, essentially "killing" its vitality (Albano, 1988)?

Ultimately, it is difficult to debate the aims and products of videogame preservation procedures because the collecting and preservation communities have so little experience with this new art form. One way to augment the discussion regarding the challenges of videogame collection and preservation is to come to a deeper understanding of the contexts and processes of its creation. The reason it is possible to speculate on Picasso and Braque's wishes or to discuss the responsibilities of the painting conservator is because there is some common understanding about the role of the "artist;" and the creative practices, materials, and methods of these more traditional artistic forms. Because we have centuries' worth of scholarship on patronage, creation, and materials selection; and we have some conception of the context of the traditional artistic creative process, it is possible have an informed debate about acceptable and unacceptable changes a preservation procedure commits to the piece. Furthermore, there is a long and proud tradition of collecting and providing access to traditional art. This is not true for videogames specifically and new media art in general. In this new model, huge teams of people with wildly divergent backgrounds come together to create videogames, working in collaborative teams consisting of visual and aural artists, producers, designers, and programmers. This conception of group artistic creation is wholly new, and itself deserves further study to understand its implications on collection, description, preservation and access. Even the videogame end product is not entirely stable: companies regularly release patches to fix programming bugs, different versions to introduce new characters and functionalities, and the quality and quantity of the players logged on at any given time make every interaction with the game different. With so many creation practices and contexts uncertain, it is not surprising that the preservation and collecting communities are facing significant challenges in the successful representation and documentation of these new media art objects. If there was a deeper comprehension of the general factors involved in new media art creation, the preservation problem might become more manageable; and if artists had a straightforward set of guidelines or "best practices" for creation, perhaps they could create more stable artifacts.

**Current Research Initiatives and Existing Collections**

There are very few research initiatives focused on videogame preservation. In addition to the currently unfunded [re]Create project described above, the Library of Congress is funding the "Preserving Creative America" project which focuses on digital film, music, and art. The project "Preserving Virtual Worlds," (McDonough, 2007) focuses on videogames – specifically immersive virtual worlds like Second Life (Linden Research Inc., 2008). The Library of Congress funding for that project is intended to study the preservation challenges inherent in collecting and preserving end-product videogames, and a major focus of that grant is to examine the necessary representation schemes, emulation and migration procedures, and to develop canonical lists of videogames for the cultural record. This funding is a milestone in the recognition of videogames as a valid and legitimate cultural form, and represents one of the only research projects focused specifically on the preservation of videogames. The work undertaken by the "Preserving Virtual Worlds" project team will be valuable for the preservation, representation, and future access to these important cultural artifacts, and will provide a useful template for future work on this subject.

Although there are few research projects devoted to videogames, there are a number of existing archives and private collections that focus on them. These run the gamut from physical archives of game hardware and software, to virtual collections of videogame music, art, and manuals (Game Preservation SIG of the IGDA, 2008). Listed below are the major collections in the United States, Europe and Japan.

### *Physical Archives Of Game Hardware and Software*

These institutional archives hold a physical collection of videogame hardware and/or software.

- Stephen M. Cabrinety Collection at Stanford University: The Cabrinety Collection on the History of Microcomputing contains commercially available computer hardware, software, realia and ephemera, and printed materials documenting the emergence of the microcomputer in the late 1970s until 1995. The collection specifically documents the emergence of computer games, with a focus on games for Atari, Commodore, Amiga, Sega, Nintendo, and Apple systems. As such, the software collection documents the increased technical ability of computer software programmers and the growing sophistication of computer-generated graphics from the early days of games like Pong to the more contemporary era of game systems like Nintendo 64. (Stanford University Libraries & Department of Special Collections, 1997)
- Ralph H. Baer Papers at the Smithsonian's Lemelson Center for the Study of Invention and Innovation: Ralph Baer was a pioneer of early videogame technology. German-born ordnance specialist, inventor, and engineer. Papers include autobiographical materials; firearms notes, manuscripts, and photographs; and videogame and television engineering notes, diagrams, schematics, and video documentation. (Smithsonian Institution, Oswald, & Fleckner, 2007)
- Computer History Museum - The mission of the Computer History Museum is to preserve and present for posterity the artifacts and stories of the information age. As such, the Museum plays a unique role in the history of the computing revolution and its worldwide impact on the human experience. While the museum collection focuses mainly on general hardware and software, it does include some game material. (Computer History Museum, 2008)
- Digital Game Archive: The DiGA e.V. was founded to establish a one-of-a-kind digital game archive on the Internet, which encourages the free download of commercial computer and videogames suitable for any platform. This Berlin-based organization provides access to nearly 30,000 games. (Digital Game Archive, 2008)
- International Arcade Museum: This collection provides access to "one of the world's leading archives covering the art, entrepreneurs, inventions, and history of the amusement and coin-operated machine industries." (International Arcade Museum, 2008) Additionally, the International Arcade Museum also maintains the KLOV, or "Killer List of Videogames;" an ever growing and comprehensive list, with related media (images and sound), of videogames.

### *Digital Game Software Repositories*

These virtual repositories hold digital archives of software, with the archive being held on servers.

- Software Preservation Society: The Software Preservation Society (SPS), formerly the Classic Amiga Preservation Society (CAPS), dedicates itself to the preservation of software for the future, namely classic games. As it is, these items are no longer available from their original suppliers, and are mainly in the possession of an ever-diminishing community of individual collectors. (Software Preservation Society, 2006)
- Archive.org Classic Software Preservation: The Internet Archive founded the Classic Software Preservation Project (CLASP) in January 2004 to help permanently archive classic, obsolete retail software from the late 1970s through the early 1990s. The Archive works to acquire copies of original consumer software of that era, and, with the help of technical partners, make perfect digital copies of these rapidly decaying floppy discs. CLASP will then lock the data away in their vaults for safekeeping, until either the copyright expires on the titles in question, or the companies who own the rights to the software make the titles freely available. (Internet Archive, 2007)

*Audio*

These online collections, run by individuals, contain game audio and music.

- Videogame Music Archive: A collection of MIDI music "from before the reign of NES through to the Battle of the NextGens." (Video Game Music Archive, 2008)
- OverClocked ReMix: This website is dedicated to arranging and re-mixing videogame music. (Overclocked Remix, 2008)
- LucasArts Soundtracks: This website collects MP3s and album art of non-Star Wars early LucasArts games. (Hofmeister, 2004)
- Slightly Dark: This website is dedicated to rare and out-of-print original videogame soundtracks and arranged albums from Japan. (Slightly Dark, 2007)
- Game Music Revolution: This website contains a game music CD information database covering mostly Japanese game music releases. (Game Music Revolution, 2006)

*Game Manuals*

This collection provides online access to videogame manuals. The physical archives listed above usually have game manual preservation as well.

- Replacementdocs: This site provides access to a large collection of high quality scanned images of instruction manuals in their full, original format with all original artwork and other graphical elements intact. (Replacement Docs, 2008)

*Strengths and Weaknesses of Currently Existing Collections*

These collections vary in scope, focus, and validity. Some are housed in large academic institutions, within the context of traditional archives; invested amateurs run others, and these collections are limited by the time and resources of that individual or group of individuals. Some of the collections focus on videogame material; others collect videogame material as a sub-set of their main institutional goal. Finally, some of these collections try to provide access to primary materials while others either provide access to end products (like video games, manuals, or individual music tracks), or they collect and organize information about end products. Many of these collections, whether housed within institutional settings or in individual server space, are managed by people who have an intense personal interest in their success; videogames are their avocation rather than their vocation, and the videogame collection isn't necessarily the main collection's focus. Due to this lack of decisive institutional support, even within institutions, videogame collections as a whole appear haphazard and fractured as opposed to collections that focus on conventional subjects, or are composed of traditional materials.

The fact that invested individuals are in charge of these collections can either be a strength or weakness depending on your point of view, but these individuals' interest, energy, and devotion are characteristics that will have a guiding influence on the development of formal institutional collections. Additionally, these somewhat ad-hoc collections set the stage for future development of more formal repositories, and provide Information professions with a valuable starting point for future collection development.

*The Videogame Archive at the University of Texas at Austin*

In an attempt to address the situation in videogame collection development, the Center for American History at the University of Texas at Austin, in collaboration with some of the leading figures in the game industry, has announced a new archive dedicated to videogames, which will be the first in Texas, and one of the few institutional archives dedicated to collecting, preserving, and making accessible those materials unique to the videogame industry. To ensure an archive of scholarly and cultural interest, the Center will gather and make available for research materials from all sectors of the industry, including developers, publishers, and artists. In addition to the games themselves, archival materials of interest include:

- Documents relating to the conception, development, planning, management, marketing, scripting, technology, and design of computer and console games.
- Art in the form of drawings, paintings, sketches, diagrams, block diagrams, play charts, environments, and other forms involved in game development projects.
- Digital files, including development documents, art, programs, source code, images, e-mail correspondence, planning data, contracts, and business plans.
- Physical game platforms including computers, consoles, cartridges, diskettes, controllers, sound boards, speakers, and especially early models that are no longer available for play.
- Collateral materials used in marketing or developing games, such as posters, cut-out figures, play weapons, photos, costumes, and vehicles used to convey the sense of the games.
- Business documents related to the operation of the game business.
- Game player material such as e-mail correspondence, Web sites, and game magazines.

By creating an institutional-level collection that focuses on all aspects of the game creation and production process, the creators of the Videogame Archive at the Center for American History at the University of Texas at Austin hope to be leaders in the field, and to attract large donations from video game pioneers and current practitioners alike. Collecting these materials will not only provide a scholarly record of videogame history, but will also enable the development of more relevant and realistic preservation models than exist today.

**Conclusion**

Massively multiplayer online video games are important and significant cultural artifacts. Not only are they worthy of meticulous and robust collection, representation, and preservation; it will increasingly become more and more important for collecting institutions to provide access to these materials. The issues involved in preservation depend on having access to primary documents relating to all aspects of the production process. Talking to videogame creators, developing models, and collecting primary production materials will support the industry, as well as facilitate the acceptance of the industry as an important cultural producer.

This paper reviewed some of the major obstacles to authentic and reliable preservation of these culturally significant new media artifacts. By reviewing the major videogame collections in the United States, Europe and Japan the current state of videogame archives and preservation procedures was revealed. These collections, while run by knowledgeable and eager individuals, are limited in their ephemerality and their focus on the end product.